\documentclass{article}

\usepackage{arxiv}

\usepackage[utf8]{inputenc} 
\usepackage[T1]{fontenc}    
\usepackage{hyperref}       
\usepackage{url}            
\usepackage{booktabs}       
\usepackage{amsfonts}       
\usepackage{nicefrac}       
\usepackage{microtype}      
\usepackage{lipsum}
\usepackage{graphicx}

\title{Designing a Mobile Social and Vocational Reintegration Assistant for Burn-out Outpatient Treatment}

\author{
  Patrick Gebhard, Tanja Schneeberger \\
  German Research Center for Artificial Intelligence\\
  Saarland Informatics Campus, Saarbr{\"u}cken, Germany \\
  \texttt{firstname.lastname@dfki.de} \\
   \And
 Michael Dietz, Elisabeth Andr{\'e} \\
  Augsburg University\\
  Augsburg, Germany\\
  \texttt{lastname@hcm-lab.de} \\
  
   \And
 Nida ul Habib Bajwa \\
  Saarland University\\
  Saarbr{\"u}cken, Germany\\
  \texttt{n.bajwa@mx.uni-saarland.de} \\
}

\begin{document}
\maketitle

\begin{abstract}
Using Social Agents as health-care assistants or trainers is one focus area of IVA research. While their use as physical health-care agents is well established, their employment in the field of psychotherapeutic care comes with daunting challenges. This paper presents our mobile Social Agent EmmA in the role of a vocational reintegration assistant for burn-out outpatient treatment. We follow a typical participatory design approach including experts and patients in order to address requirements from both sides. Since the success of such treatments is related to a patients emotion regulation capabilities, we employ a real-time social signal interpretation together with a computational simulation of emotion regulation that influences the agent's social behavior as well as the situational selection of verbal treatment strategies. Overall, our interdisciplinary approach enables a novel integrative concept for Social Agents as assistants for burn-out patients.
\end{abstract}


\section{Introduction}




Mental health-care systems and the treatment of patients come with high standards. Mental disorders are highly prevalent and emotionally demanding for individuals, families, and society. From affected persons, many individuals remain untreated although effective treatments exist. This so-called treatment gap differs between different mental disorders but is generally high, e.g., 56,3\% (median, worldwide) for major depression \cite{kohn2004treatment}. For severe cases like a burn-out, where a clinical stay is required, not only the primary care lacks coverage, but also the outpatient treatment afterward. 

Interactive Social Agents are, due to the incorporation of social skills (e.g., active listening, mimicry, gestures, emotion models) \cite{bickmore2005establishing, gratch2007creating, gratch2013using, gebhardetal2018}, becoming more and more able to develop a relationship with people. Since this ability is important to tackle mental grievances \cite{davis1994nonverbal}, such agents could play a supportive role in technologically supported treatment.

What is commonly known as a burn-out, would be usually classified as a \textit{major depression} and belongs to the class of affective disorders. One explanation for the occurrence of these disorders is an emotion regulation dysfunction \cite{beck1979cognitive}. Thus, a Social Agent designed to assist the social and vocational reintegration for burn-out patients after a clinical stay needs a model of emotions, emotion regulation, and associated sequences social signals to simulate those in an affective user model. 

This paper presents a design for a Social Agent in the role of an assistant for burn-out patients in outpatient treatment. The design has been approved by the external ethic counseling board of the project. After presenting related work, we will elaborate on the requirements of patients and therapists focusing on general characteristics, the interaction design, and the provided content of the reintegration assistant. Due to the sensitivity of the domain and the possible social impact, all ethical, legal, and social implications are discussed with an external ethic counseling board. Then, the paper focuses on needed software components. All privacy-related patient data is handled without using cloud-based solutions. In the end, we will sketch a way of introducing the assistant in the patients' life and our first planned evaluation. Both rely on the results of discussions with the ethical counseling board about how to successfully introduce such a technological approach in order to support both the patients' and the therapists' requirements. 


\section{Related Work}
The health context as a use case for Social Agents has been getting attention in research for about 15 years. 

The Fit Track with the Relational Agent Laura is one of the first systems \cite{bickmore2005establishing}. Laura has the role of an exercise advisor that interacts with patients for one month on a daily basis to motivate them to exercise more. Laura was equipped with different effective patient-provider communication skills (e.g., empathy, social dialogue, nonverbal immediacy behaviors) to build and maintain good working relationships over multiple interactions. A study showed that the use of those relational behaviors significantly increases working alliance and desire to continue working with the system. This work suggests that computer systems that interact with patients, especially those that engage patients in dialogue or long-term, repeated interactions, can benefit by explicitly designing in emotional and relational communication behavior. 

Lucas et al. \cite{lucas2014s} showed which positive effects autonomous Social Agents could have in a health-care setting to overcome the barrier to receiving truthful patient information. Patients have to disclose sensitive information to their practitioners where honesty is an essential factor. However, out of fear of self-disclosure and the feeling of being judged negatively by the practitioners \cite{farber2003self}, patients often hold back information. To overcome this problem, the authors compared two different interviewers in a health-screening interview \cite{lucas2014s}. Participants interacted with a virtual human and were led to believe that the virtual human was controlled by either humans or automation. Compared to those who believed they were interacting with a human operator, participants who believed they were interacting with a computer reported lower fear of self-disclosure and were rated by observers as more willing to disclose.

The potential use of virtual humans as counselors in psychotherapeutic situations was investigated by Kang and Gratch \cite{kang2010virtual}. Assuming that self-disclosure of the patient is crucial in psychotherapy, they are examining in \cite{kang2010virtual} with which conversational partner participants disclose more private information measured with self as well as with external assessment. Their study reveals that a virtual human can elicit more self-disclosure in a hypothetical conversational scenario than a human in a raw or degraded video. 
The outcomes of their study suggest a possibility of using virtual humans as counselors in psychotherapeutic situations because virtual humans can provide high anonymity, maintaining communicators' privacy when they reveal intimate information about themselves. However, in this study, the participant was in the role of an interviewee being asked questions to find out if she is a suitable match with whom an apartment could be shared, which is not the same like being in a psychotherapeutic setting. 

Regarding the design of stress management agents, Martin et al. \cite{martin2018} discuss underlying requirements and challenges including the critical need to support both personalization and conversation. They propose i.a. that the system needs knowledge about the user's preferences and capabilities regarding remedial activities. 

Generally, it seems that the exploitation of Social Agents in health-care might have unique possibilities and even advantages to support the health-care system to overcome existing challenges.



\section{Background and Concepts}

\subsection{Burn-out and Typical Care}
\label{sec:Burnout-Care}
Burnout is not classified as a mental disorder according to the latest version of the International Classification of Diseases (ICD-11). However, most burnout patients are diagnosed with a \textit{Major Depression}, an affective disorder which is related to dysfunctional emotion regulation strategies \cite{beck1979cognitive}. In cognitive behavior therapy, patients who are suffering a severe form of depression are admitted to a clinic and offered medical and psychotherapeutic assistance to help cure their psychological disorder. In such cases, there is often a need for stabilizing patients for weeks or months before they can be discharged from the hospital and sent for outpatient treatment \cite{cuijpers2011psychological}. A long stay at the hospital is necessary to create a safe space for patients, where patients do not feel the burden of having to accomplish anything. The clinical stay is merely a phase of stabilization such that patients improve sufficiently and can be sent back home into their typical environment. Very often, patients sent home for outpatient treatment relapse as the safe space created in the hospital environment is completely missing \cite{vittengl2007reducing}. The long-term success of depression treatment severely relies on a seamless outpatient treatment which incorporates a guided re-integration into work \cite{niehaus2008betriebliches}. However, this is often difficult in countries such as Germany as there are huge waiting lists outpatient therapy \cite{priebe2006provision}. Moreover, outpatient therapy differs significantly from a prolonged stay at the hospital, where ample opportunities exist to talk to nurses, doctors, and psychotherapists, giving patients a sense of security \cite{schneider2005adherence}. 

\subsection{Social Agents as Reintegration Assistants}
Social Agents in the role of reintegration assistants have the task to facilitate the reintegration into work for people who have undergone stabilizing psychotherapy for depression in a clinic for several weeks. As an always present and accessible assistant, such agents provide guidance, diary for self-monitoring, and coaching, whenever needed. With the patient's agreement, the system can provide information (e.g., specific incidents at the workplace, difficulties at home) to the patient's therapist.

After leaving the clinic, patients face different challenges in everyday life and at work. Both environments come with socio-emotional challenges. At home, patients have to master different tasks, e.g., washing clothes, cooking or picking up children while continuing a regular daily routine. Besides, they have to be alert for depression symptoms and act accordingly. At work, persons concerned have to talk to their superiors and colleagues to find a model for the reintegration process. Usually, those models include the identification of a "healthy" workload and a step-by-step increase of working hours that has to be monitored. Depending on their position, they probably have to find and to accept a new role in the organization. Socio-emotional-wise, patients might have to face 1) that they feel incompetent in general, 2) that they feel helpless to tackle all the challenges on their own in an "insecure" environment or 3) that they are considered to be cured what might be opposed to their actual perception. Without or with little protecting environment, they have to build self-efficacy and to recognize personal limits. In order to ward off the danger of loneliness, it is helpful to expand social resources and (re-)integrate into family and friends and (re-)learn to rely on them. For all those tasks, an reintegration assistant has to support the user with, e.g., 
tracking, exercises or psycho-educational information.

\subsection{Ethical, Legal, and Social Implications}
\label{ELSI}
Due to its personal nature and impact on the society in general, the creation and research of reintegration assistants has to be accompanied by a board that monitors the Ethical, Legal, and Social Implications (\textit{ELSI board}) with experts in each area. They provide recommendations and approve the design of the assistant.

\section{Design of EmmA}
\label{sec:design}
The creation of the EmmA application follows a participatory design approach. This includes several expert interviews of therapists (schools: cognitive-behavior therapy and cognitive psychoanalysis) and patients. Those interviews revealed crucial aspects that we take into account for both the application design and the holistic introduction of the app to the users. Against the initial planning, the clinical experts opted for introducing the app already during the patient's stay in the clinic. They argue that this holistic approach should improve the commitment of the users. On the other hand, also the therapists get acquainted with the application. This can be helpful because therapists can also use the application during outpatient psychotherapy. During the first period of cognitive-behavior therapy, patients usually have one session with the therapist per week while there is no contact between patient and therapist in the meantime. In the time between the sessions, patients could use the application to monitor their behavioral change and practice new skills. This could improve the emphasized between-session change \cite{goldfried2003therapist} in cognitive-behavior therapy.

We identified four classes: general characteristics, relational agent behavior, interaction scenarios, and application services.

\subsection{General Characteristics}
Self-disclosure of the patient is known to be highly relevant since the first days of psychotherapy. Patients' self-disclosure is affected by aspects of the therapist, the interaction with him, the patient's appraisal of the subject he is talking about, and aspects of the patient himself \cite{farber2003patient}. Also, for our presented application, it is essential that users self-disclose when interacting with the Social Agent. The amount of self-disclosure of the user is dependent 
on the virtual agent, e.g., her trustworthiness \cite{wheeless1977measurement} or her own self-disclosure \cite{kang2011people}. Therefore, EmmA is designed in a way that "she" fosters users' self-disclosure by addressing the user empathically.


\subsection{Relational Agent Behavior}
\label{sec:agent-behavior}
Although the presented concept is not intended to replace therapy, it is crucial for its success that some characteristics of a successful patient-therapist-relationship are applied. In this light, we favor a natural interaction via voice with the agent in general. A silent text chat option is offered (Fig.~\ref{fig:Lydia}, chat symbol).

Nonverbal behavior, such as facial expression, gaze, body movement, and gesture play an important role in patient-therapist interactions \cite{davis1994nonverbal}. Therefore, the nonverbal behavior of the assistant might be just as crucial. In therapy, several positive nonverbal behaviors could be identified. They include a moderate amount of head nodding and smiling; frequent, but not staring, eye contact; active, but not extreme, facial responsiveness; and a warm, relaxed, interested vocal tone \cite{hall1995nonverbal}. To have similar positive effects, we will design the agent to show adequate back-channeling behavior while the user is talking (e.g., nodding or smiling). Moreover, the agent will have a general affirmative and active aura to affect the user positively, which is communicated by a moderate amount of head nodding and smiling while speaking. The combination of the possibility to subtly manipulate the agent's facial expressions and the real-time social signal interpretation enables us to realize a (re-)active, but not extreme, facial responsiveness. Moreover, we will enliven the speaking parts of the assistant with suitable facial expressions and upper body movements. 

Version 1 aus Template (keine Ahnung, warum es nicht geht. Ist doch eigentlich ein einfacher Befehl):

\begin{figure} 
    \centering
    \includegraphics{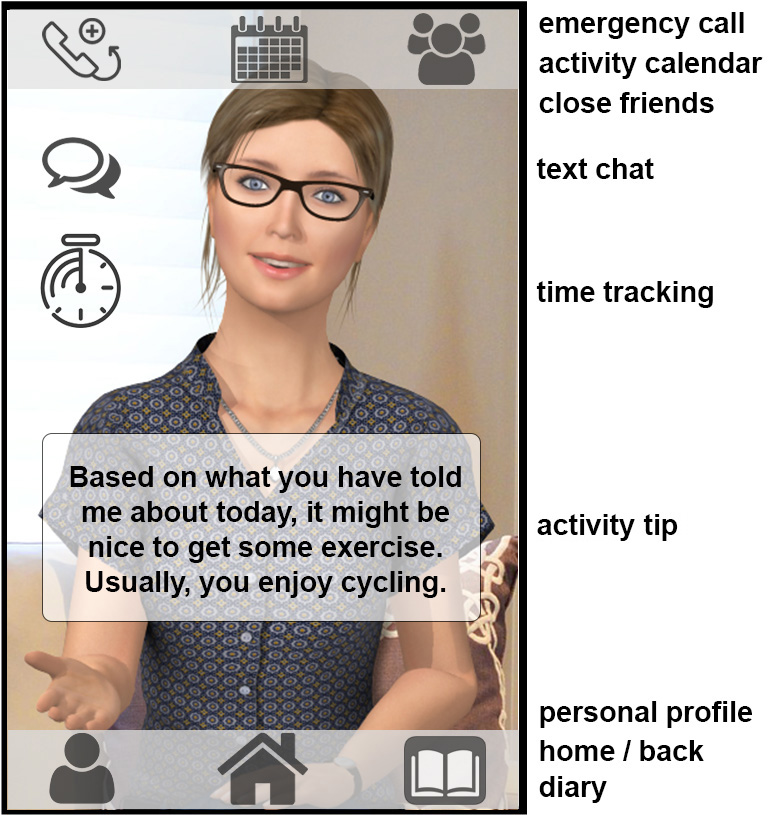}
      \caption{Application screen concept with service buttons.}
  \label{fig:Lydia}
\end{figure}

Mimicry describes the phenomenon that people unconsciously and automatically imitate other people in interactions and social situations. 
For the psychotherapeutic process, mimicry seems to have three major roles. First, it has been proposed as a therapists' tool to improve their understanding of their patients. From this point of view, mimicry is an active process to generate empathy, i.e., an understanding of the patient's feelings and perspective. Second, therapists' use of nonverbal behavior mimicry seems so have a positive effect on patients' evaluation of the therapist as well as the therapeutical relationship. Third, mimicry does not only improve the subjective assessment of the patient but also his objectively measured mental health condition~\cite{hess1999mimicry}. The mimicry behavior of the Social Agent EmmA follows the consideration that communicated affect expressions serve both intrapersonal and interpersonal emotion regulation \cite{baenninger-huber96, magai95}. They relate to the process of grounding \cite{clark96} and attribution, which has been investigated in studies by Bavelas et al. \cite{bavelasetal00}. Therefore, EmmA does not "copy" the complete nonverbal behavior of the user, since it might lead to 1) mimicry of negative emotions or 2) an intensification of a user's negative self-attribution. As a general rule, EmmA will only mimic positive social signals like smiles if there is a low chance of simulated negative internal emotions (Sec. \ref{MARSSI}). Note, negative internal emotions might be reflected by positive communicated emotions (e.g., smiling because of a feeling of insecurity). Hence, EmmA, e.g., would show a neutral non-verbal mimicry behavior even if the patient appears fine by communicating happiness (e.g., smile), if there is a chance that the patient is experiencing a negative internal emotion (e.g., shame). In all other cases social signals will be mimicked with a neutral expression (e.g., nodding instead of smiling). All mechanisms are employed in order to positively motivate patients, as human therapists would do. If these mechanisms work as good as they work for human therapists will be part of the evaluation.


\subsection{Interaction Scenarios}
\label{InteractionScenario}
EmmA provides three interaction scenarios: 1) \textit{first acquaintance}, 2) \textit{daily features}, and 3) \textit{weekly features}. The goal of the first acquaintance is that the user gets to know the reintegration agent and builds trust what is crucial for a successful work relation. Moreover, information about the user is gathered in both questionnaires and conversations with the assistant. To build a user model, the assistant asks, e.g., questions about health history or gives a personality questionnaire. Following the goal-setting theory \cite{locke1994goal}, one of the most important parts of the cognitive therapy \cite{margraf2008lehrbuch}, the user sets personal goals regarding the working time after the professional re-entry and defines psychological resources (e.g., favorite activities) together with the agent. In the daily interaction, the assistant will track these goals, remind the user of these and adapt them if needed (Fig.~\ref{fig:Lydia}, stopwatch). In the last step of the first acquaintance, the assistant explains the features of the app during a walk-through. During the daily interaction, EmmA is aware of expressed emotions and the user model, especially regarding the personal emotion regulation strategies, is refined and validated with each interaction. These include conversations about critical situations of the past day or about future events that might invoke negative emotions (Fig.~\ref{fig:Lydia}, calendar, diary). Another daily feature is the assessment of personal variables, such as drive, strain, sleep and monitoring of the well-being in the form of rating scales (Fig.~\ref{fig:Lydia}, personal icon). The weekly functions comprise a mood barometer and a working hours graph. The former is based on the weekly course of the saved emotions and personal well-being. It allows the user to see on first glance how he/she has felt during the week. The graph is therefore divided into emotions, drive, and strain. After presenting the mood barometer, the agent asks the user to reflect on some special occasions. This information is recorded, transcribed and made visible as a diary entry. The working hours graph compares the actual working hours with the aspired ones for each day of the last week. This graph serves for inspection of the goal attainment. In case the goal was not achieved, the assistant asks again to reflect on this, which will also be recorded in the diary.

To sum up, the application offers a self-assessment as well as elicits concrete measures for improving the health of the user at the beginning. Additionally, the application regularly brings the user to reflect on his/her well-being by asking the questions mentioned above. All those features serve as assistance during the outpatient-therapy after initial inpatient treatment.


\subsection{Application Services}
\label{Application}\label{sec:PychoEducation}
The EmmA app provides several features based on research on depression, which can be divided into three categories: 1) \textit{acquisition of the patient profile}, 2) \textit{psychoeducation}, and 3) \textit{stimulating action}: 

1) The patient profile (Fig.~\ref{fig:Lydia}, personal icon) is acquired both spoken and automatically. The Social Agent EmmA will ask users to provide goals for monitoring purposes (e.g., working hours, personal and life goals) as well as a list of activities which they find positive. The goal-setting is supposed to be one of the most important parts of the treatment of depression patients \cite{margraf2008lehrbuch}. Thus, following the goal-setting theory of Locke and Latham \cite{locke1994goal}, the users will enter important and concrete goals. EmmA will ask users twice a day (in the morning and in the afternoon) how their day went and what they did. These regular interactions will be used to assess various elements related to the success of reintegration: Emotional states will be assessed via smartphone sensors, drive (i.e., the level of physical activity) will be assessed by the accelerometer and time spent on work will be assessed through manual check-ins and check-outs. 

2) With regard to their \textit{psychoeducation}, many outpatients lack time to reflect on their stress level and resources during the week and are asked to create a daily diary detailing stressful events and their thoughts \cite{margraf2008lehrbuch}. Using sensory and speech data collected by the app, a mood graph will be generated at the end of the week and will provide a novel and easier way of reflecting success and challenges during the week. Lastly, in an e-Learning section users can learn about, e.g., typical behaviors during depressive episodes, possible actions against a lack of drive, possible methods to improve sleep quality, warning signs of overload. Moreover, this section will include classical psychoeducational information. Psychoeducation refers to the process of providing education and information to the patient. The goal is to enable a healthy lifestyle by imparting knowledge and skills. It improves the understanding of a disease and its treatment by rethinking and correcting erroneous ideas about their disorder through new, scientifically based knowledge \cite{pitschel2017psychoedukation}. Originally developed for schizophrenic patients~\cite{anderson1986schizophrenia}, it is now applied for various mental disorders \cite{bauml2015handbuch}. For patients with Major Affective Disorders, psychoeducation has been associated with a better resolution of the index episode and better patient global outcome \cite{glick1994effectiveness}.

3) As a self-learning app, EmmA will gather information over the course of usage and provide helpful tips on actions that can be taken. After having gathered sufficient data, the app will proactively guide users on doing positive activities if, e.g., a lack of drive was noticed (i.e., not having moved much during the day) or questions will be asked to assist users to take time and ponder about successes and challenges during the week (Fig.~\ref{fig:Lydia}, activity tip).


\let\originalparagraph\paragraph
    \renewcommand{\paragraph}[2][.]{\originalparagraph{#2#1}}

\section{Technology}

\begin{figure} 
    \centering
     \includegraphics[width=1.0\textwidth]{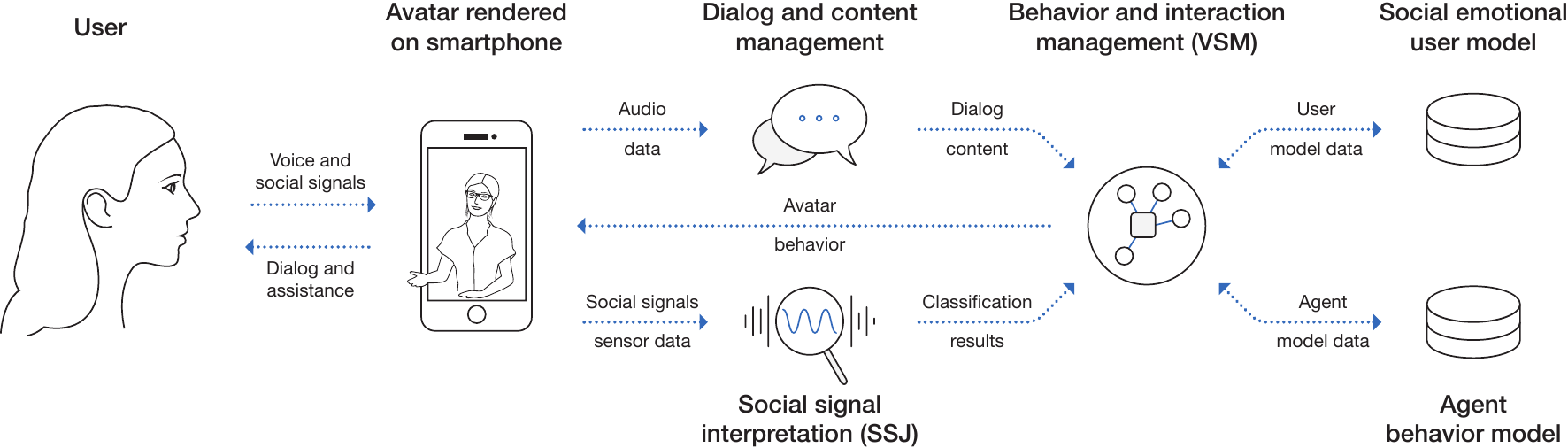}
      \caption{Architecture of the vocational reintegration assistant EmmA}
  \label{fig:achitecture}
\end{figure}

EmmA consists of several components: 1) WebGL Social Agent rendering with TTS unit\footnote{\url{https://vuppetmaster.de}}, 2) Nuance-based on-device NLU, dialog, and content management\footnote{\url{https://www.semvox.de/en/technologies/odp-india}}, 3) real-time social signal interpretation framework (SSJ), as well as 4) a behavior and interaction modeling and execution tool (VSM) using a social-emotional user model and an agent behavior model (Fig. \ref{fig:achitecture}). All components are implemented in JAVA\textsuperscript{TM} for the Android ($\geq$ API v. 16) ecosystem. For the sake of this paper's topic, we focus on the explanation of the 3\textsuperscript{rd} and 4\textsuperscript{th} components features.

\subsection{Mobile Social Signal Interpretation}
\label{sec:ssj}
In order to capture the user's emotional and physiological cues we make use of the mobile social signal interpretation framework SSJ\footnote{\url{https://hcm-lab.de/ssj/}}~\cite{damian2018,damian2016}. The framework enables the recording, analysis, and recognition of human behavior based on social signals such as gestures, facial expressions, body movements, and emotional speech. For that, SSJ allows to interface with and extract data from device internal and external sensors. Due to its modular architecture, the data processing in SSJ is performed through pipelines, which consist of a sequence of autonomous components that allow the parallel and synchronized signal processing. Additionally, SSJ supports machine learning pipelines for the execution of pre-trained models as well as on-device training of simple online learning classifiers such as Na\"{\i}ve Bayes. This is especially useful for creating machine learning models that can be adapted to the individual user behavior over time, which is one of the main goals of the present work.

\let\originalparagraph\paragraph
    \renewcommand{\paragraph}[2][.]{\originalparagraph{#2#1}}

While the computational power of mobile devices has rapidly increased in the past few years to a point where mobile social signal processing is feasible, training complex classifiers such as neural networks directly on these devices within a reasonable amount of time is still a challenging task and mostly limited by the currently available hardware. One approach to bypass this problem is to use transfer learning. Thereby an existing model for a specific domain is retrained to solve a classification problem in another domain~\cite{pan2010}. Since modifying an existing model requires fewer resources than training a new one from scratch, the current generation of mobile devices is already capable of performing this task, as shown by Seiderer et al.~\cite{seiderer2018}. In their work, they used existing deep neural network models for image classification and only retrained the last layer directly on the device. 
Although other approaches propose cloud-based solutions where data is transferred to powerful servers for model training~\cite{eshratifar2018}, we only focus on techniques where the data does not leave the user's device. The main reason for this is that in our intended application scenario the model data can contain highly sensitive information about a user's personal life such as habits, behaviors, and relations, which need to be protected to increase the user's trust towards the system. For the EmmA application we therefore only consider online and transfer learning approaches which can be trained directly on the user's personal device.

In order to enable the system to react to the user's behavior appropriately, the system first needs to identify the mental and emotional state of the user. For that, previous research in the area of emotion and stress detection primarily focused on physiological signals such as electrocardiography~\cite{vrijkotte2000}, skin conductivity~\cite{hernandez2011}, or pulse~\cite{zhai2005}. Additionally, eye movements~\cite{distasi2013} and pupil dilation~\cite{mokhayeri2011} were examined as indicators of stress in controlled environments. More recent work investigated the use of mobile sensors in naturalistic environments to record relevant data for stress detection with the aim of gaining clues about everyday stressors from target users. Preliminary studies by Sano and Picard~\cite{sano2013} showed that people who are exposed to high levels of stress can be reliably identified by their level of activity and their mobile phone usage - even when compared to self-reports. 
%
Based on these results, we will initially analyze the data from the internal smartphone sensors such as camera, microphone, accelerometer, and gyroscope. Depending on the results, additional external sensors such as smart watches or activity trackers will also be considered for the collection of physiological data and social cues. Through the analysis of these cues, we hope to gain insights into the user's current state of mind which will be used to create a Social Agent that intelligently interacts with and reacts to the user. This allows us to observe the impact of the agent on the user and enables us to evaluate the effectiveness of different approaches for model personalization and adaption.

\subsection{User Model and Agent Behavior Model}
\label{MARSSI}
The interactive agent EmmA assists burn-out patients in their vocational reintegration. In most cases, burn-out symptoms (e.g., exhaustion, withdrawal, feeling of emptiness) can be related to dysfunctional emotion regulation (Sec. \ref{sec:Burnout-Care}). In this light, a computational model of user affect is employed. It simulates internal emotion regulation strategies that are connected to observable sequences of social signals. This model extends the MARSSI computational model of appraisal, regulation, and social signals \cite{gebhardetal2018} by 1) validating existing classifiers for emotion regulation social signal sequences and 2) incorporating new signal sequence classifiers that are explicitly created to recognize emotion regulation signal sequences from burn-out patients. MARSSI uses an emotion notation differentiating \textit{communicative emotions} from \textit{situational emotions} and \textit{structural emotions}. The last two are internal emotions reflecting individual subjective experiences. Structural emotions (e.g., shame as a highly prevalent emotion in therapy) represent information about the appraisal of oneself, hence connected to the self-image. Situational emotions represent information that is linked to specific situations and reflect the level of security. Both emotion classes are relevant for modeling internal affective states of burn-out patients. Communicative emotions are defined as observable social signal sequences and are connected to internal emotions by regulation strategies. MARSSI provides four signal classifiers that are related to the four emotion regulation strategies described by Nathanson \cite{nathanson94} in the context of experiencing the structural emotion shame: 1) Avoidance, 2) Attack Self, 3) Attack Other, and 4) Withdrawal. These are represented as cognitive rules and used to simulate parallel plausible appraisals and emotions, their regulation, and regulating emotions. Using this model, it is, e.g., explainable that a smile, embedded in a stream of specific other social signals, will lead to a simulation of the structural emotion shame with a particular value of chance (calculated by the relevant classifier). Next to individual patterns of emotion regulation, personality aspects will be taken into account for refining the social-emotional user model. Such aspects are collected in the initial interaction phase as well during continuous interaction (Sec. \ref{Application}). 

All simulated emotions will be attached as situational emotions to situation representations (cf. Dialog and content management, Fig.~\ref{fig:achitecture}) of daily activities (e.g., debriefing of the daily work experience, Sec.~\ref{sec:TypicalInteractionwithEmmA}) that a patient exchanges with EmmA. If a patient revisits such a situation, it is assumed that s/he has an affective bias that reflects the connoted situational emotions. By the real-time observation of the patient's signals these will be corroborated or not. Based on this information, EmmA is then able to choose between different strategies how to tackle a difficult situation. For example, if it is recognized that the goal of leaving work at an early time (as part of the reintegration strategy) is connoted with shame, EmmA chooses relevant supporting strategies, which are defined by experts. Because EmmA is able to identify difficult emotional situations, the patient could discuss these, by giving access to the patient's therapist to EmmA's analysis. Since MARSSI provides a cognitive layer that can be exploited to generate explanations on the possible simulated emotions, EmmA can explain what has been assumed emotional-wise. This simulation of patients emotions, in fact, is used as an example of the application's build-in psychoeducation about emotions (Sec.~\ref{sec:PychoEducation}). Patient's feedback is used to refine the individual emotion simulation.

EmmA's non-verbal interaction behavior, esp. mimicry behavior, is influenced by the simulation of possible internal emotions. If the model simulates negative internal emotions with a high value of chance, albeit a patient would show positive expression of emotions (e.g., smile), the agent would show a neutral non-verbal mimicry behavior (Sec.~\ref{sec:agent-behavior}).

\subsection{Behavior and Interaction Management}
EmmA's behavior and interaction are controlled by a hybrid agent behavior model, which is executed in real-time on the device. It combines scripted and autonomous behavior strategies. The latter covers methods of mimicking social signals of the user. This mimicry model is able to learn parameters of specific user behavior (e.g., dynamics of nodding) in order to provide a "known" (to the specific user) interaction behavior if needed. Apart from the agent behavior model, agent's real-time behavior depends on 1) the current dialog context, 2) simulated emotional state of the user, and 3) current recognized social signals (Sec.~\ref{sec:agent-behavior}, Sec~\ref{sec:ssj}, and Sec.~\ref{MARSSI}). The behavior model is maintained by VisualSceneMaker\footnote{http://scenemaker.dfki.de} (\textit{VSM})~\cite{gebhardetal2012}. It is specifically designed to meet the requirements for modeling interpersonal coordination in social human-agent interaction \cite{mehlmannetal2016}.

\section{Typical usage of EmmA}
\label{sec:TypicalInteractionwithEmmA}

As described (Sec.~\ref{InteractionScenario}), there are different phases of the interaction. Upon the first usage, data protection declarations have to be accepted. Here, users are informed about data collection and processing. After that, the acquaintance phase starts in which the user gets to know the reintegration assistant EmmA and vice versa. EmmA shows up on the screen for the first time, introduces herself, asks for the users' name and explains the versatility of the app as well as its main goal. After this introduction, the assistant assesses the actual status of the user via a short interview. During this interview, EmmA asks the user, for example, how long his/her career break was, whether he/she was only on sick leave, in rehab and whether he/she is now in outpatient therapy to gain enough information of the previous psychological health history of the user. Furthermore, the diagnosis (including ICD-11-diagnosis code) and support (e.g., contact details from friends, therapist) are collected. In the last step of the first use protocol, the assistant explains which features are accessible across the app, and a walk-through is offered. The available buttons are home, E-learning, personal profile, close friends, activity calendar, emergency call, time tracking and text chat (Fig. \ref{fig:Lydia}). When clicking the home button, the user returns to the assistant. The button E-learning leads to information material and positive activities. Personal data and the diary are available via the profile button and contact details via the friends' button. The calendar button offers the user the possibility to save appointments and reminders. The emergency call button provides the user with emergency numbers. The time tracking button is used to log in and out working hours and thereby assess the working time. Finally, the text chat button allows the user to chat with EmmA via text messages instead of talking to it. After having completed the first use scenario, users are able to use the daily features.

\begin{figure}[h] 
    \centering
    \includegraphics[width=1.0\textwidth]{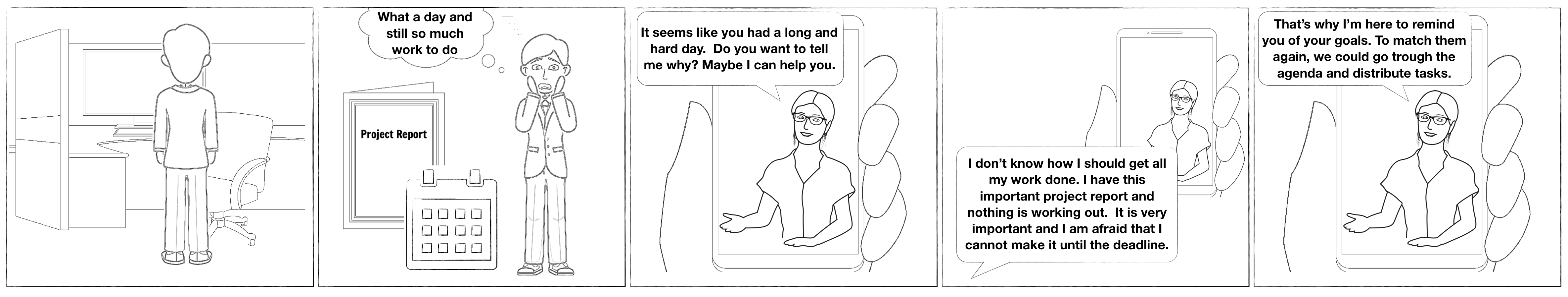}
      \caption{Storyboard excerpt showing a challenging situation that the reintegration assistant is trying to solve with the user.}
  \label{fig:Story}
\end{figure}

During the daily interaction the app tracks, e.g., the actual working hours. In order to achieve this, the user has to check-in in the morning when he/she starts working and check-out when the user finishes work. This function is used to assess the attainment in regards to the working hours. It is important to mention that the assistant explains that this information about the working hours is not sent to third parties. The only purpose is for the user to get an overview of his/her working hours. The actual working hours are compared to the aspired ones, and in case of non-compliance the reintegration assistant would address this in an interaction in the afternoon or evening of the same day (Fig.~\ref{fig:Story}). EmmA talks to the user about the recent emotions and gives advice about activities to pursue if negative emotions are dominant. Another daily feature is the assessment of personal variables, such as drive, strain, sleep, and monitoring of the well-being. The assistant asks the user to rate on a scale on 1-7 how rested, how energetic and how strained the user feels. This information is later used for the weekly features. The assessment of emotions as well as the assessment of personal variables can be performed several times a day, but takes place at least two times a day, in the morning and the evening. 

Weekly, EmmA will recap the last week with the user. The graphs created from the collected data, allow the user to see on first glance how he has felt during the week. After presenting this graph, EmmA asks the user to reflect on some special occasions, e.g., "Why did you feel so bad on Thursday morning?". The answer is recorded, transcribed and made visible as a diary entry. 

On a general level, interactions between the user and the assistant can be triggered by both of them. Whenever the user wants to interact with the assistant or get some information, he can open the application. On the other hand, at least two times a day, the app reminds the user via push-messages to talk to the assistant or give some information needed.

\section{First Evaluation}








All presented design aspects (Sec.~\ref{sec:design}) are approved by an external ELSI board. One of the topics discussed in greater length with them includes, e.g., the data protection issues with the conclusion that we have to ensure complete confidentiality and trust by the user the data is processed only on the device itself.

We plan a first evaluation of the app as soon as we have finished the implementation of the demo version. The overall goal of the first evaluation is to find out if our reintegration assistant matches the requirements of patients and therapists. In a qualitative field experiment, we will evaluate the app for one week with patients and therapists in burnout outpatient therapy. At the beginning of the week, users will have an initial meeting with the experimenter, including questionnaires about, e.g., attachment style and personality. Moreover, they will have a walkthrough of the application given by the experimenter. Possible questions can be cleared and the planned procedure of the trial will be explained. During the week, users should interact with the system and provide feedback about their interaction with it at the end of each day. After one week, there will be a final interview with the users. All the gathered data of the interactions between user and assistant will be analyzed to find out, e.g., which problems users address to the system. Additionally, we will acquire data that is needed for improving the natural language understanding component.

An ethical approval for the mentioned studies within the project has already been obtained from both our external ELSI board and the ethical review board for mathematics and computer science.


\section{Summary and Future Work}
This paper presents a technology-driven concept for complementing the current state-of-the-art of burnout outpatient treatment. Core aspects are the employment of the mobile Social Agent EmmA in the role of a caring assistant. EmmA is explicitly designed to fill the existing support gap after the successful clinical stabilization phase in which burn-out patients have to (re-)adapt to every-day life and work-life demands. Usually, such patients have one outpatient therapy session per week.

EmmA is an always present burnout outpatient reintegration assistant offering desired and relevant services that are identified by patients and therapist experts. The most relevant service areas are supporting daily and weekly tasks, psychoeducation, and stimulating actions. If the patient agrees, the therapist can discuss difficult emotional (e.g., shame eliciting) situations that EmmA is able to identify by observing the patient's emotion regulation capabilities in interactions. For communicating empathically, EmmA employs a real-time social signal interpretation together with a computational simulation of emotion regulation. Based on this, a real-time dynamic behavior model adapts the agent's behavior to the current social-emotional situation as well as the situational selection of verbal support strategies, which are defined by experts. Both emulate important non-verbal and verbal support strategies of therapists. The goal is to build a relationship between patient and agent in order to activate trust and self-disclose, which are essential factors. Technology-wise, the system relies, besides the mentioned components, on a realistic rendering of the agent, a real-time voice recognition, and natural language understanding. All personal data and data processing components are on-board and are entirely controlled by the patient. An external ELSI board has approved the design and technological approach.

Our evaluation plan of the next years addresses several aspects of the system. Most pressing is the evaluation if this approach leads to a higher success rate of reintegration for burn-out patients with fewer patients relapsing. Another part covers the assessment of various aspects concerning the acceptance of the system and its parts. Within that scope, we manipulate the agent systematically and its behavior to find out what is needed to enhance the self-disclosure of burnout patients like suggested by, e.g., Kang and Gratch \cite{kang2010virtual}. Together with the external ELSI board, we discuss 1) the possibility to integrate an automatic detection of depression alarm signals, which would trigger interventions by the patient's therapist, 2) the adaption of this approach for other domains of anxiety disorders such as social phobia.

\section{Acknowledgements}
This work is partially funded by the German Ministry of Education and Research (BMBF) within the EmmA project (funding code 16SV8029) and the German Research Foundation (DFG) within the DEEP project (funding code 392401413). We thank the Charamel GmbH for realizing our requirements with regard to the virtual agent environment.

\bibliographystyle{unsrt}  
\bibliography{references}  


%



\end{document}